\documentclass{IEEEtran}
\usepackage[pdftex]{graphicx,xcolor}
\usepackage[fleqn]{amsmath}
\usepackage{newtxtext}
\usepackage[varg]{newtxmath}

\usepackage{amsfonts, amssymb,bm}

\usepackage{algorithm,algorithmic}
\usepackage{enumerate,ascmac,here,comment,wrapfig,cite}

\usepackage{bm}
\usepackage{mathrsfs}
\newcommand{\1}{\mbox{1}\hspace{-0.25em}\mbox{l}}
\newcommand{\defeq}{\overset{\bigtriangleup}{=}}

\newtheorem{theorem}{Theorem}
\newtheorem{lemma}{Lemma}

\author{
  Yutaka Jitsumatsu, 
  Ukyo Michiwaki, 
  and 
  Yasutada Oohama,
 \thanks{ A part of this paper was presented in the 41st Symposium on Information Theory and Its Applications (SITA2018). }
 \thanks{Y. Jitsumatsu and U. Michiwaki are with Kyushu University, Japan.}
 \thanks{Y. Oohama is with the University of Electro-Communication, Japan.}
}

\title{The Conditional Information Leakage Given 
Eavesdropper's Received Signals in Wiretap Channels}

\begin{document}
\maketitle
\begin{abstract}
Information leakage in Wyner's wiretap channel model 
is usually defined as the 
mutual information between the secret message and the 
eavesdropper's received signal. 
We define a new quantity called 
\lq\lq conditional information leakage given the 
eavesdropper's received signals,'' which expresses
the amount of information that eavesdropper gains 
from his/her received signal.
A benefit of introducing this quantity is that we can develop a fast
algorithm for computing the conditional information leakage, which
has linear complexity in the code length $n$, while
the complexity for computing the usual information leakage is exponential in $n$. 
Validity of such a conditional information leakage as a security criterion 
is confirmed by studying the cases of binary symmetric channels and binary erasure channels. 


\end{abstract}

\section{Introduction}

Information-theoretic security is a concept that we design a cryptosystem 
so that our private information must be kept hidden even if 
adversary's computational power is unlimited, as opposed to 
cryptography schemes whose secureness critically depends 
on computational hardness assumption~\cite{LiangNowPublisher}.
Such an information theoretically secure systems include
secret sharing~\cite{Shamir1979}, 
private information retrieval~\cite{Private_Information_Retrieval}, and 
Wyner's wiretap channel~\cite{Wyner75a}.

Wiretap channel is a model of physical layer security
in wireless communications, 
in which Alice wants to transmit a positive-rate message 
to Bob reliably and securely  
in the presence of the eavesdropper Eve, where
the coding scheme is open to Eve as well as Bob. 
A discrete memoryless wiretap channel is described by a conditional 
probability $p(y,z|x)$, where $x$, $y$, and $z$ denote
the symbols for input of the channel, output of the main channel (the channel from Alice to Bob),
and the output of eavesdropper's channel (the channel from Alice to Eve). 
Wyner~\cite{Wyner75a} proved that there exists a sequence of encoding and decoding systems 
in which Alice can securely transmit a message to Bob while
keeping the information leakage to Eve arbitrarily small if the coding rate $R$
is smaller than the secrecy capacity defined by 
  \begin{align}
    C_S = \max_{p_X} \left\{ I(X;Y) - I(X;Z) \right\}, \label{secrecy_capacity}
  \end{align}
where $X$, $Y$ and $Z$ are random variables for $x,y$ and $z$, 
$I(X;Y)$ is the mutual information between $X$ and $Y$, 
and $p_X$ is a probability mass function for $X$. 

After Wyner's pioneering work, many research has been done. 
Wyner studied a degraded wiretap channel, where $X\to Y\to Z$ forms a Markov chain,
while Csisz\'ar and Korner extended the channel model to 
a broadcast channel with confidential messages~\cite{Csiszar-Korner78}.
The secrecy condition that Wyner posed was that $\frac1n I(S; Z^n) \to 0$ 
as $n\to \infty$, where $S$ denotes the random variable of a secret message
and $Z^n = Z_1, Z_2,\ldots Z_n$ is a random vector of Eve's received symbols.
Maurer and Wolf~\cite{Maurer-Wolf} pointed out that such a secrecy criterion is weak
and they posed a stronger constraint that $I(S; Z^n) \to 0$ 
as $n\to \infty$. They showed that the secrecy capacity (\ref{secrecy_capacity})
can be achieved even under the strong secrecy criterion. 
Hayashi derived a secrecy exponent~\cite{Hayashi2006} 
which shows that there exists a sequence of encoders 
by which the information leakage goes to zero exponentially. 

Construction of wiretap channel codes has been extensively studied. 
Coset-coding, also referred to as syndrome-coding, is generally used. 
A wiretap channel codes using low density parity check (LDPC) 
codes~\cite{Thangaraj} and that using polar 
codes~\cite{Mahdavifar2011,Csacsouglu2013} were studied. 
Codes in~\cite{Thangaraj, Mahdavifar2011} satisfy weak secrecy, while
the wiretap codes based on polar codes~\cite{Csacsouglu2013} satisfy strong secrecy, i.e., codes in~\cite{Csacsouglu2013} is proved to 
satisfy that unnormalized information leakage 
goes to zero as $n$ goes to infinity. 

The purpose of our research is the evaluation of the secureness of a given encoder.
Our fundamental question is that when $n$ is finite, how much information is leaked to Eve 
for an explicitly designed wiretap code. 
This situation is the same as the evaluation of bit error rate, i.e., 
we must evaluate the bit error rate of an explicitly designed code, 
even if it is a capacity-achieving code.  
However, there is a difficulty in evaluation of information leakage; 
it takes exponential time in $n$ for computing 
the mutual information $I(S^m; Z^n)$. 
This fact comes from the assumption that the eavesdropper is allowed 
to access to unlimited computational resources. 
Thus, previous studies for evaluating the information leakage 
focus on special cases where 
the eavesdropper's channel is a 
binary erasure channel (BEC) or a binary symmetric channel (BSC). 
Ozarow and Wyner~\cite{OzarowWyner} considered the case where the main channel is noiseless and Eve's channel is a BEC\footnote{
To be precise, Eve is assumed to access $\mu$ symbols 
out of $n$ transmitted symbols of her own choice~\cite{OzarowWyner}. 
Such a model is called a type II wiretap channel. }. 
In this case, information leakage is obtained by computing 
the rank of submatrix consisting of all $j$-th rows of the parity check matrix,
where $j$'s are the indices of the erased bits. 
Zhang et al.~\cite{Zhang,Al-Hassan} 
restricted their attention to the case 
when Eve's channel is a BSC and main channel is noiseless. 
Probability generating function is used to 
efficiently compute the information leakage. 
Unfortunately, Zhang et al.'s method 
cannot be applied for other DMCs to compute the information leakage. 
Mori and Ogawa evaluated an upper and lower bound of 
the information leakage for binary symmetric wiretap channel 
and reduced the complexity for computing the
information leakage~\cite{MoriOgawa2018}. 

In this paper, we provide a method for evaluating the amount 
of information leakage to Eve 
when the Alice's encoding system is a coset coding 
and eavesdropper's channel is a general binary-input discrete 
memoryless channel (BI-DMC). 
The contribution of this paper is four-fold.
\begin{enumerate}
\item[1)] We introduce a new quantity called 
"conditional" information leakage given 
Eve's received signal, defined by 
$L(z^n) =  \mathscr{H}(S) - \mathscr{H}(S|Z^n=z^n)$,
where $\mathscr{H}(X)$ denotes the entropy of a random variable $X$. 
Introducing such a new quantity is the key of this paper. 
The standard definition of the information leakage 
$I(S; Z^n)$ is the expectation of $L(Z^n)$ over $Z^n$.

\item[2)] We propose a method for computing $L(z^n)$ when coset coding is employed and
Eve's channel is a BI-DMC. 
The proposed method is a modified version of Zhang et al.'s method~\cite{Zhang, Al-Hassan}. 
Extension of the proposed method to $M$-ary input $M\geq 2$ is straightforward. 

\item[3)] We show that if Eve's channel is a BEC in addition to the condition stated in 2), 
then $L(z^n)$ is equal to the number of bits on $S$ that Eve gains from $z^n$. 
This fact supports that the definition of $L(z^n)$ is reasonable. 
We also show that $L(z^n)$ is computed by the rank of a submatrix of the
parity check matrix $A$ for the coset coding. 

\item[4)] We show that, under the conditions 2) and 3), 
the probability distribution of $L(z^n) = L(z^n|A) $ which depends on $A$ 
can be well approximated by the ensemble average of the probability distribution over random $A$. 
\end{enumerate}

The rest part of the paper is organized as follows. 
In Section 2, we give definitions of the wiretap channel model and 
the coset coding.
In Section 3, we define the "conditional" information leakage 
and give an efficient method for computing it when Eve's channel is a BI-DMC. 
Computation of the conditional information leakage when Eve's channel is a BEC
is also discussed. 
In Section 4, the probability distribution of the conditional 
information leakage when Eve's channel is a BEC is discussed. 
The ensemble average of the probability distribution
over a all possible coset coding is given. 
Section 5 concludes this paper.

\begin{figure}[t]
  \includegraphics[width=8cm]{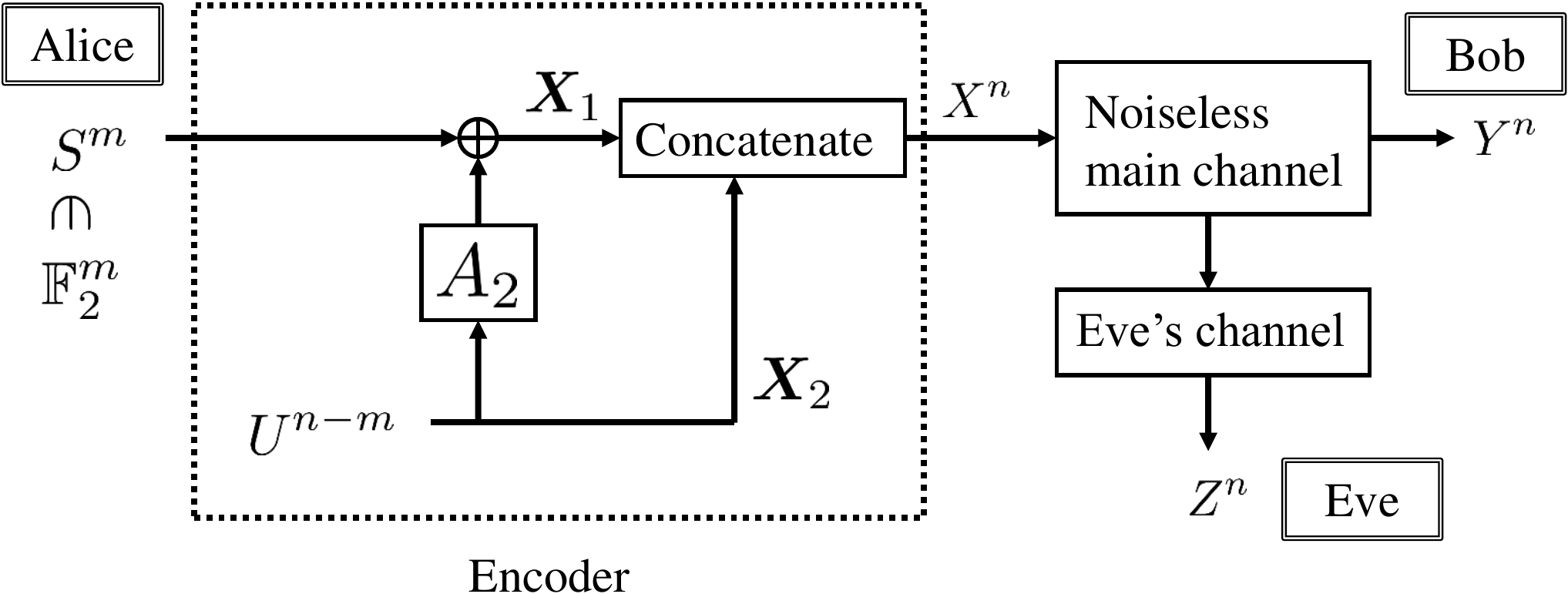}
  \caption{System model: A wiretap channel with noiseless main channel. Coset coding with a parity check matrix $A \in \mathbb{F}_2^{m\times n}$ is employed as a encoder. $S^m$ is Alice's secret message and $U^{n-m}$ is a random bit sequence.}
  \label{fig:wtc}
\end{figure}

\section{Wiretap Channels with Coset Codes}
  \label{cosetsection}

Consider a wiretap channel with noiseless main channel
and a coset coding 
as the encoder (See Fig.\ref{fig:wtc}).
Alphabets of the input and Bob's received symbols are $\mathbb{F}_2$,
and that of Eve's received symbol is arbitrary, denoted by $\mathcal{Z}$.
In Section 2 and 3, we consider BSC and BEC for Eve's channel. 

Let $A \in \mathbb{F}_2^{m\times n}$ ($m<n$) be a parity check matrix, where $\mathbb{F}_2$ is a finite field of order $2$. 
Assume that $A$ is constructed as $A=[I_m \  A_2]$, where $I_m$ is
the identity matrix of $m$-th order. As in \cite{Zhang,Al-Hassan,MoriOgawa2018}, we assume that the main channel is noiseless.
Let $S^m \in \mathbb{F}_2^{m}$ be a column vector of $m$ random variables for the secret message.
For a fixed $S^m=s^m$, a codeword is randomly chosen from $ C(s^m) = \{ x^n| A x^n = s^m\} $,
where $C(s^m)$ is the coset with a coset leader $s^m$. Then, the codeword 
for a secret message $s^m$ is represented by
a random variable $X^n$ that follows a uniform distribution on $C(s^m)$. 

Then, the codeword can be expressed by using 
$S^m \in \mathbb F^m_2$ and a column vector of $n-m$ uniform random variables 
$U^{n-m} \in \mathbb F^{n-m}_2$ as 
    \begin{align}
      X^n = 
      \begin{bmatrix}
        \bm{X}_1\\
        \bm{X}_2
      \end{bmatrix}
       =
      \begin{bmatrix}
        S^m\\
        0
      \end{bmatrix}
        + 
      \begin{bmatrix}
        A_2 \\
        I_{n-m}
      \end{bmatrix} U^{n-m}
    \label{Coset_Coding}
    \end{align}
It is easy to check that $S^m = A X^n$ holds. The assumption of the noiseless main channel makes 
Bob's received signal $Y^n$ equal to $X^n$, and therefore Bob can  perfectly recover $S^m$ by computing $A Y^n$. 


\section{Conditional information leakage given Eve's received signal}
  \subsection{Definition}
    Wyner defined the information leakage by the mutual information between 
Eve's received signal $Z^n$ and the secret message $S^m$, denoted by $I(S^m;Z^n)$. 
$\frac1n I(S^m; Z^n)$ or $I(S^m; Z^n)$ have been used 
as the security measure~\cite{Csiszar-Korner78, Maurer-Wolf, Hayashi2006,
Thangaraj,Mahdavifar2011,Csacsouglu2013,Zhang,Al-Hassan,MoriOgawa2018}.
Let us revisit this measure. 
Before receiving $Z^n$, Eve does not have any knowledge about the 
transmitted message $S^m$. Thus, 
her best guess for $S^m$ is that $S^m$ is uniformly distributed. 
After Eve receiving $Z^n=z^n$, where $z^n$ is a realization of $Z^n$,
her best guess for $S^m$ is that $S^m$ follows 
the a posteriori probability $p_{S^m|Z^n}(s^m|z^n)$.
Equivocation for $S^m$ is reduced from $\mathscr{H}(S^m)$ to $\mathscr{H}(S^m|Z_n = z_n)$. 
Thus, we can define the amount of information  on $S^m$ that Eve gained by
receiving $z^n$ is 
\begin{eqnarray}
   L(z^n) = \mathscr{H}(S^m) - \mathscr{H}(S^m|Z^n=z^n).\label{newI0}
\end{eqnarray}
We see that the mutual information $I(S^m;Z^n)$ is the 
expectation of Eq.(\ref{newI0}) with respect to $Z^n$.
The use of $I(S^m;Z^n)$ as the security criterion is reasonable since 
the code designer does not know the realization of $Z^n$ beforehand.
However, when we perform a computer simulation, $z^n$ is available and
therefore we can treat $L(z^n)$ as the information leakage. 
We refer to $I(S^m;Z^n)$ as the average information leakage and
$L(z^n)$ as the conditional information leakage given $z^n$. 
Here we give a remark that Eq.(\ref{newI0}) can take negative value in general.
However, if {\it a prior} probability distribution $P_{S^m}$ is uniform,
then $\mathscr{H}(S^m) = m$ and therefore Eq.(\ref{newI0}) is always nonnegative.

    \begin{figure}[t]
    \begin{center}
      \includegraphics[width=5cm]{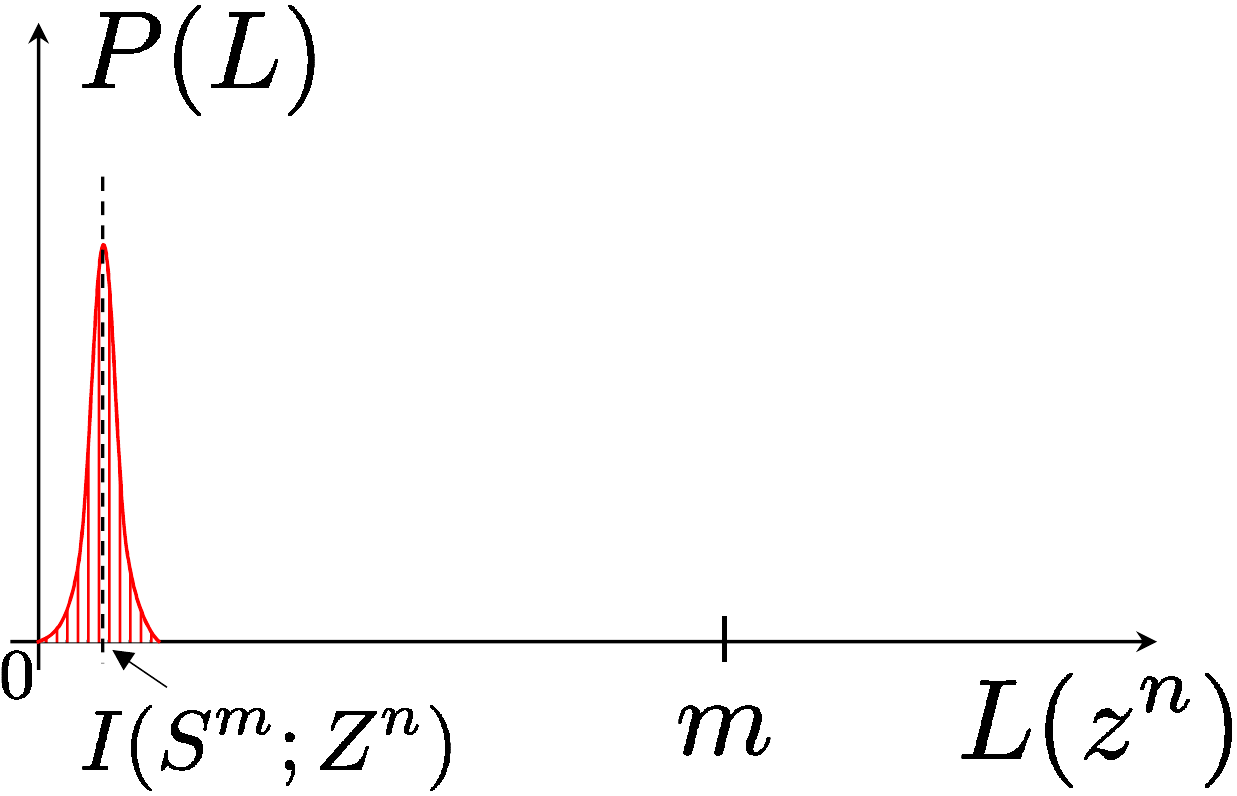}
      \caption{Illustration of the distribution of "conditional" information leakage given $z^n$.}
      \label{fig:one}
    \end{center}
    \end{figure}

It is considered that $L(Z^n)$ is distributed according to the probability distribution of 
$Z^n$. As the eavesdropper's received signals are random variables, 
it is natural to consider information leakage is also a random variable.
An image of the distribution of $L(Z^n)$ is illustrated in Fig.\ref{fig:one}. 
The distribution of $L(Z^n)$ is more informative than the average information
leakage $I(S^m;Z^n)$ which is a single scalar and is equal to the mean value of $L(Z^n)$.

\subsection{Computation of the conditional information leakage}

\subsubsection{Probability Generating Function}
In~\cite{Zhang,Al-Hassan}, Zhang et al. proposed an efficient 
method for computing the information  leakage when the encoder is a coset code with parity check matrix 
$A \in \mathbb{F}^{m \times n}$ and wiretapper's channel is a BSC.
They used the probability generation function for computing the probability 
distribution of $A V^n$, where $V^n$ denotes the random vector expressing
bit flip in the BSC. 
Note that Zhang et al.'s method is only applicable when Eve's channel is a BSC. 

For a random vector $ \bm{X} = (X_1,X_2,...,X_m)^T \in \mathbb F_2^m$, 
where $(\cdot)^T$ denotes the transpose of a vector or a matrix, 
we define the probability generation function $G_{ \bm{X} }(t)$ as 
    \begin{align}
      G_{ \bm{X} }(t) 
   = E[ t^{ \bm{X} } ] 
   = \sum_{ \bm{x} \in \mathbb{F}_m^2 } 
     P( \bm{X} = \bm{x} ) t^{\bm{x}}. 
    \end{align}
The following property of the probability generating function is useful. 
Consider the "exclusive or" of two independent random vectors 
$X$ and $Y \in \mathbb{F}_2^m$ denoted by 
$\bm{X} \oplus \bm{Y} = (X_1 \oplus Y_1, \ldots, X_m \oplus Y_m)^T$. 
The probability generation function of $X\oplus Y$, 
$ G_{ \bm{X} \oplus \bm{Y} }(t) = E[ t^{ \bm{X} \oplus \bm{Y} }]$ is computed by 
    \begin{align}
    G_{ \bm{X} \oplus \bm{Y} }(t) 
& = G_{ \bm{X} }(t) G_{ \bm{Y} }(t) \label{temp31}. 
    \end{align}
The conditional probability mass function $P_{S^m|Z^n}(\cdot|z^n)$
can be computed using the property (\ref{temp31}).

  \subsubsection{Computation of average information leakage for BSC: The conventional method}
  \label{BSCL}
In this subsection, we explain Zhang et al.'s method \cite{Zhang, Al-Hassan} for computing 
the information leakage of a wiretap channel 
under the assumption that main channel is noiseless, and 
the eavesdropper's channel is a BSC with
crossover probability $\delta$ and
the coset code with a parity check matrix $A$ is employed. 
Zhang et al.~\cite{Zhang,Al-Hassan} proved that $\mathscr{H}(S^m| A Z^n) = \mathscr{H} ( A V^n) $ holds
and  
gave an efficient computation method for $ \mathscr{H} ( A V^n) $, by which
mutual information $I(S^m ; A Z^n) = \mathscr{H}(S^m) - \mathscr{H}(S^m|A Z^n)$ is computed. 
In~\cite{Zhang,Al-Hassan}, relation between the information leakage $I(S^m ; Z^n)$
and $I(S^m ; A Z^n)$ was not explicitly given. 
In order to clarify the relation, we give the following theorem: 

   \begin{theorem} 
    \label{eqtheo}
    Consider a wiretap channel with noiseless main channel. 
    Assume that Eve's channel is a BSC and 
    the encoder uses a coset code with a parity check matrix $A =[ I_m | A_2 ] 
    \in \mathbb{F}_2^{m \times n}$. 
    Let $V^n$ be a random vector expressing the noise of the BSC, 
    i.e., $V_i=1$ if $i$-th bit is flipped and $V_i=0$ otherwise. 
    Then we have, 
      \begin{align}
        I(S^m; Z^n) = m - \mathscr{H} (A V^n).  
      \end{align}
    \end{theorem}

This theorem shows that $I ( S^m ; Z^n) = I( S^m ; A Z^n)$ holds.
We give a proof of this theorem in~\ref{appendix.A}. 

A naive computation of the probability distribution of $A V^n$ requires us 
to take a time proportional to $2^n$. 
Zhang et al. showed that the computational complexity can be reduced by the use of
probability generating function. Zhang et al. gave the following theorem~\cite{Zhang,Al-Hassan}:
    \begin{theorem}[\cite{Zhang,Al-Hassan}]
      \label{Zhangtheo}
      Consider a wiretap channel with noiseless main channel.
      Suppose that the eavesdropper's channel is a BSC with crossover probability
      $\delta$ and a coset code with a parity check matrix 
      $A = [ \bm a_1\ \bm a_2 \cdots \bm a_n]$ is used. 
      The BSC is expressed by an additive noise vector $V^n = V_1 V_2 \cdots V_n$. 
      Then, the probability generating function of $ A V^n $ is 
      \begin{align}
      \label{temp7}
        G_{A V^n}(t) = \prod_{i=1}^n \left((1-\delta) + \delta t^{ \bm{a}_i }\right). 
      \end{align}
    \end{theorem}
Although the proof was given in~\cite{Zhang}, to make this paper self-contained, 
we give a proof here. 

{\it Proof:} We have 
      \begin{align}
        A v^n &= \bm a_1 v_1\oplus \bm a_2 v_2 \oplus 
                 \cdots \oplus \bm a_n v_n \label{temp21}.
      \end{align}
Therefore, the following chain of equalities hold. 
      \begin{align}
        G_{A V^n}(t)&= \sum_{ \bm{x} \in \mathbb{F}_2^m } {\rm Pr}( A V^n  = \bm{x} ) t^{\bm{x}} \nonumber\\
                   &= \sum_{v^n \in \mathbb F_2^n} p(v^n) t^{ A v^n }\nonumber\\
                   &\overset{(a)}{=} \prod_{i=1}^n \sum_{v_i \in \mathbb F_2}p(v_i) t^{ \bm a_i v_i}\nonumber\\
                   &= \prod_{i=1}^n \left((1-\delta) + \delta t^{ \bm{a}_i }\right)\label{temp12}. 
      \end{align}
Step (a) follows from the equality in (\ref{temp21}) together with the assumption that 
$V_1,V_2,...,V_n$ are independent. \hfill$\square$

    By Theorem \ref{Zhangtheo}, the probability generation function of $A V^n$ is expressed by 
    the multiplication of $n$ terms. Expansion of  
    Eq.(\ref{temp12}) as a polynomial of $t$ is expressed as  
    \begin{align}
      G_{A V^n}(t) = \sum_{ \bm{x} \in \mathbb{F}_2^m} \beta_{\bm{x}} t^{\bm{x}}, 
      \label{GAVn-1}
    \end{align}
    where $\beta_{\bm{x}}$ is the probability of the event $A V^n=\bm{x}$. 
    We can compute $\beta_{\bm{x}}$ recursively, as follows:
    For a given $A = [\bm a_1 \ \bm a_2\ ...\ \bm a_n]$, we define for $r=1,2, \ldots, n$, 
    \begin{align}
      G_{A V^n}^{(r)}(t) = \prod_{i=1}^r \left( (1 - \delta) + \delta t^{ \bm{a}_i } \right). 
    \end{align}
    Let the expansion of $G_{A V^n}^{(r)}(t)$ be 
    $G_{A V^n}^{(r)}(t) = \sum_{ \bm{x} \in \mathbb{F}_2^m } \beta_{\bm{x}}^{(r)} t^{\bm{x}}$. 
    Then, we have 
    \begin{align}
       G_{A V^n}^{(r+1)}(t) 
    &= G_{A V^n}^{(r)}(t) 
       \left( (1 - \delta) + \delta t^{ \bm{a}_{r+1} } 
       \right) 
    \nonumber\\
    &= (1-\delta) G_{A V^n}^{(r)}(t) + \delta \sum_{\bm{x} \in \mathbb{F}_2^m } 
       \beta_{\bm{x}}^{(r)} t^{ \bm{x} \oplus  \bm{a}_{r+1} }
    \nonumber\\
    &= \sum_{\bm{x} \in \mathbb{F}_2^m} \left\{ (1-\delta) \beta_{\bm{x}}^{(r)} 
       + \delta \beta^{(r)}_{ \bm{x} \oplus \bm{a}_{r+1} }\right\} 
       t^{ \bm{x} }, 
      \label{GAVn-2}
    \end{align}
      and hence 
    \begin{align}
      \beta_{ \bm{x} }^{(r+1)} 
    = (1-\delta) \beta_{\bm{x}}^{(r)} + \delta \beta^{(r)}_{ \bm{x} \oplus \bm{a}_{r+1} }
    \label{temp22}. 
    \end{align}
    Eq.(\ref{temp22}) shows that we can compute $\beta_{\bm{x}} = \beta_{\bm{x}}^{(n)}$ 
    recursively with initial value 
    \begin{align*}
    \beta_{ \bm{x} }^{(0)} = 
    \begin{cases}
    1,  \bm{x} = (0,\ldots 0)^T\\
    0,  \text{otherwise}. 
    \end{cases}
    \end{align*}
    Hence, the computation time is linear in $n$, while it is still exponential in $m$
    because of the computation of $\beta^{(r)}_{ \bm{x} \oplus \bm{a}_{r+1} }$ in (\ref{temp22}). 
    Further reduction of computational time, for example using Mori and Ogawa's method\cite{MoriOgawa2018},
    is left to be studied in future.



  \subsubsection{Computation of the conditional information leakage for BI-DMCs: The proposed method}
  \label{common}
    In this section, we give
a method for computing the conditional information leakage $L(z^n)$.
    We have 
    \begin{align}
      L(z^n) &= \mathscr{H}(S^m) - \mathscr{H}(S^m|Z^n=z^n)\nonumber\\
                   &= m + \sum_{s^m \in \mathbb F_2^m} p(s^m|z^n)\log p(s^m|z^n). 
    \end{align}
    Therefore, we compute the conditional probability distribution of $p(s^m|z^n)$ for a given $z^n$. 
    We define the conditional probability of backward channel as 
    \begin{align*}
      \Phi(x|z) \defeq \frac{P_{X}(x)W_E(z|x)}{\sum_{x' \in \mathcal X}P_X(x')W_E(z|x')} = \frac{W_E(z|x)}{\sum_{x'}W_E(z|x)},
    \end{align*}
    where $W_E(z|x)$ denotes the conditional probability for Eve's channel. 
    The second equality follows from 
the assumption that $s^m$ and $u^{n-m}$ follow uniform distributions. 
We give the following theorem:

\begin{theorem} \label{Theorem_DMC} 
Let $\beta_{\bm{s}}$ be the probability of the event $ S^m = \bm{s}$ given $Z^n = z^n$.
Then, $\beta_{\bm{s}} = \beta_{\bm{s}}^{(n)}$ is obtained by computing 
\begin{align}
  \beta_{\bm{s}}^{ (r+1) } =
  \Phi(0|z_i) \beta_{\bm{s}}^{(r)} 
+ \Phi(1|z_i) \beta_{\bm{s} \oplus \bm{a}_{r+1} }^{(r)}
\label{Eq.Theorem_DMC}
\end{align}
for $r=0, 1, \ldots, n-1$ with initial values
$\beta^{(0)}_0 = 1$ and $\beta^{(0)}_j = 0$ for $j \neq 0$. 
\end{theorem}

We give a proof of Theorem~\ref{Theorem_DMC} in~\ref{Appendix:Theorem_DMC}

By Theorem ~\ref{Theorem_DMC}, 
we can compute $p(s^m|z^n)$ by a modified Zhang et al's method, 
which is obtained by 
simply
replacing $(1-\delta)$ and $\delta$ in (\ref{temp22}) with $\Phi(0|z_i)$ and $\Phi(1|z_i)$,
respectively. 

  \subsection{Computation of the conditional information leakage when Eve's channel is a BEC}
  In this section, binary erasure channel (BEC) is assumed for Eve's channel.
  Suppose that the main channel is noiseless and coset coding is used. 
  We show that, unlike the BSC case, the conditional information leakage $L(Z^n)$ is distributed. 
  We compute $L(z^n)$ for a BEC as follows:
    \begin{theorem}
    \label{Theorem_BEC}
   Suppose that Eve's channel is a BEC and that Eve receives $Z^n = z^n$.
  Let $A = [\bm a_1\ {\bm a_2}\ ...\ {\bm a_n}]$ be the parity check matrix for 
the coset code. 
   Then the information leakage to Eve is 
      \begin{align}
        L(z^n) &= m - {\rm rank} [ \bm{a}_j : z_j = \mathrm{e} ], 
        \label{def:L}
      \end{align}
	where $ [ \bm{a}_j : z_j = \mathrm{e} ] $ denotes the submatrix of $A$ consisting
        of all $\bm{a}_j$'s for which $z_j=\mathrm{e}$. 
    \end{theorem}

This is almost the same statement as~\cite[Lemma 4]{OzarowWyner}.
However, the authors in~\cite{OzarowWyner} assumed  
a combinatorial variation of an erasure channel
in which Eve observes $\mu$ symbols 
out of the $n$ transmitted symbols.
Thus, in order to use the proof of Lemma 4 in~\cite{OzarowWyner}
as the proof of Theorem~\ref{Theorem_BEC},
a translation is needed. 
Therefore, we give a direct proof of 
Theorem~\ref{Theorem_BEC} in~\ref{proof_theorem_BEC}.

It should be noted $L(z^n)$ is independent of the erasure probability $\delta$.
Theorem~\ref{Theorem_BEC} shows that $L(z^n)$ is equal to the number of
bit on $S^m$ that Eve can recover from $z^n$. 
This is explained by the following example:

\textbf{Example 1}: 
Let $m=2, n=3$ and 
$$A = [\bm{a}_1 \ \bm{a}_2 \ \bm{a}_3 ] =
\begin{bmatrix}
1 & 0 & 1\\
0 & 1 & 1
\end{bmatrix}. 
$$
The secrete message is denoted by $(S_1, S_2)$. 
By (\ref{Coset_Coding}), the codeword
is expressed by $X_1 = S_1 \oplus U$, $X_2 = S_2 \oplus U$ and $X_3 = U$,
where $U\in\mathbb{F}_2$ is a binary uniform random variable.

If no bit is erased so that $Z_i  = X_i$ for $i=1,2,3$,
the secret message $(S_1, S_2)$ is leaked to Eve.
Suppose one of $Z_i$s is erased. 
If $Z_1 =\textrm{e}$, Eve does not know $X_1$ but she obtains $X_2$ from $Z_2 \oplus Z_3$. 
If $Z_2 =\textrm{e}$, Eve does not know $X_2$ but she obtains $X_1$ from $Z_1 \oplus Z_3$. 
If $Z_3 =\textrm{e}$, Eve does not know $X_1$ or $X_2$ but she obtains $X_1\oplus X_2$ from $Z_1 \oplus Z_2$. 
Thus, the amount of information leakage is exactly $1$ bit if one of $Z_i$s is erased. 
Suppose two of $Z_i$s are erased. If $Z_2$ and $Z_3$ are erased, Eve cannot extract
any useful information from $Z_1 = S_1 \oplus U$. 

We can also confirm that the probability distribution of $S^m$ given $Z^n=z^n$
is obtained by using the probability generating function.
For example, for the case of $(Z_1, Z_2, Z_3) =  (1, 0, \mathrm{e})$ 
we have $G_{S^m|Z^n=z^n} = ( 0 + 1 t^{\bm{a}_1} ) (1 + 0 t^{\bm{a}_2}) (\frac12 + \frac12 t^{\bm{a}_3})
= \frac12 t^{ \bm{a}_1 }+ \frac12 t^{ \bm{a}_1 \oplus \bm{a}_3}$.
Since $\bm{a}_1 = (1, 0)^T$ 
and $\bm{a}_1 \oplus \bm{a}_3 = (0, 1)^T$, we have 
\begin{align*}
\mathrm{P}( (S_1, S_2 ) = (1,0) ) &= \mathrm{P}( (S_1, S_2 ) = (0, 1) ) = \frac12\\
\mathrm{P}( (S_1, S_2 ) = (0,0) ) &= \mathrm{P}( (S_1, S_2 ) = (1, 1) ) = 0. 
\end{align*}
Then, we have $\mathscr{H}(S^m | Z^n = z^n) = 1$, i.e., remaining ambiguity on $S^m$ is exactly 1 bit.
Therefore the conditional information leakage is $I(S^m ; Z^n = z^n) = m - \mathscr{H}(S^m | Z^n = z^n) = 1$.

\section{Distribution of the conditional information leakage}
  \subsection{The case that Eve's channel is a BSC}
  \label{section:BSC} 
  In the coset coding, both $s^m$ and $u^{n-m}$ are 
  uniformly distributed, which makes $x^n$ also uniformly distributed. 
  Thus, in the case of BSC, we have 
  $\Phi(x|z)=1-\delta$ if $x=z$ and 
  $\Phi(x|z)=\delta$ if $x\neq z$. 
  Then, the probability distribution obtained by (\ref{Eq.Theorem_DMC}) 
  depends on $z^n$. 
  However, we have the following:

  \begin{theorem} \label{Theorem_conditional}
  If Eve's channel is a BSC, for any $z^n \in \mathbb{F}_2^n$, 
  the conditional information leakage is 
  independent of $z^n$, i.e., we have 
  \begin{align*}
    L(z^n) & = I(S^m;Z^n=z^n) \\
           & = I(S^m;Z^n) = m - \mathscr{H}(A V^n). 
  \end{align*} 
  \end{theorem}
  We give the proof in~\ref{Appendix:Theorem_conditional}

Theorem~\ref{Theorem_conditional} shows that $L(Z^n)$ takes $I(S^m;Z^n)$ with probability one. 
BSC is a special case where $L(Z^n)$ is a fixed value. 
For a general DMC, $L(Z^n)$ is distributed, as shown in 
the next subsection.

\subsection{The case that Eve's channel is a BEC}
\begin{figure}[t]
\centering 
\includegraphics[height=50mm]{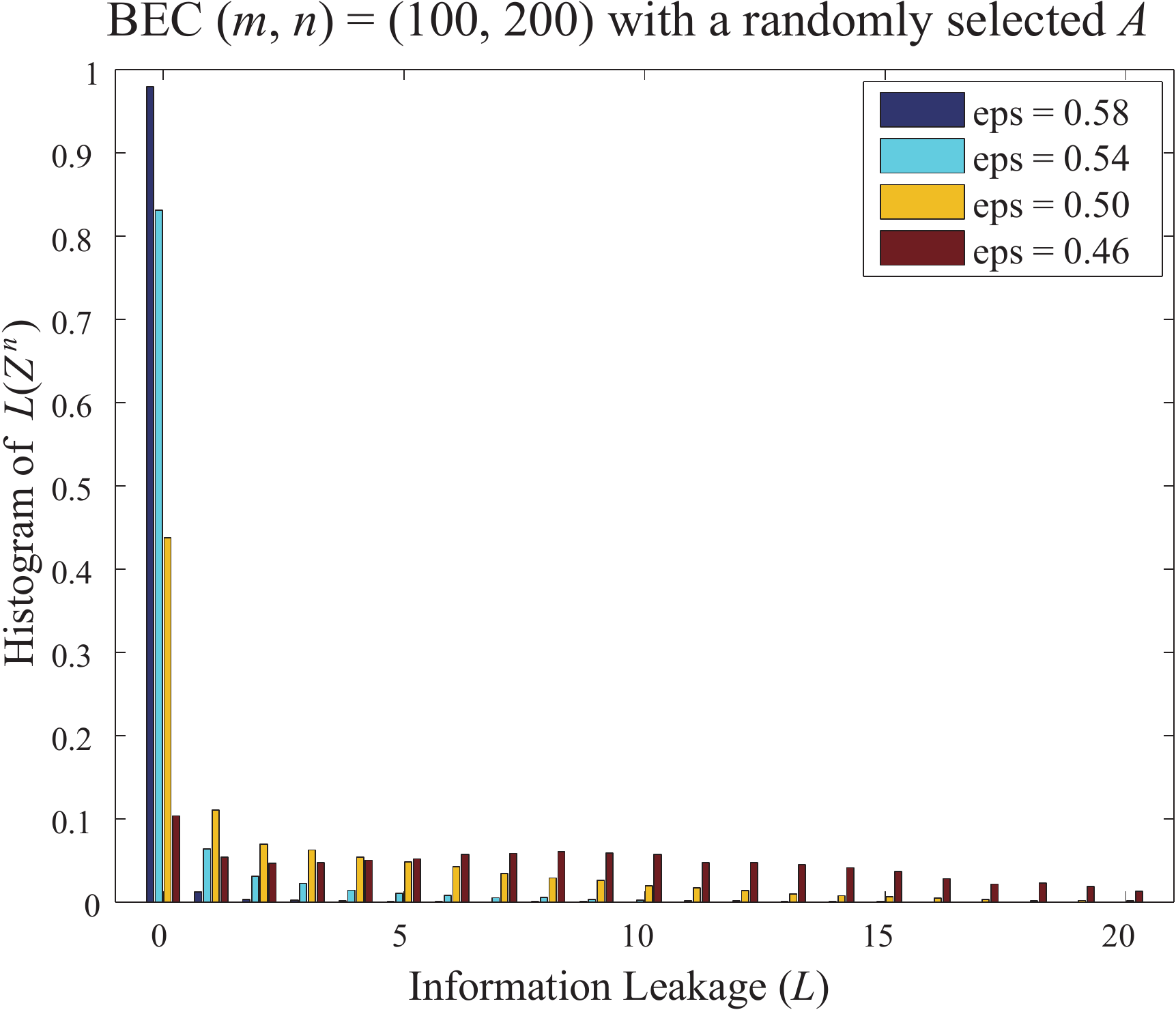}
\caption{The histogram of $L(Z^n)$ for a fixed parity check matrix $A$, 
where the number of samples is $10^4$}
\label{Distribution_of_Information_Leakage_Given_H}
\end{figure}

If Eve's channel is a BEC, 
the conditional information leakage $L(z^n)$ 
is given by (\ref{def:L}). Thus it depends on
the pattern of erased bits in $z^n$. 
To obtain the distribution of $L(z^n)$ exactly for a given 
parity check matrix $A$, we have to compute (\ref{def:L})
for all $z^n$, taking a time proportional to $2^n$.
This subsection provides a numerical result by Monte Carlo simulation.

In the simulation, we generate $z^n$s, say $z^n(1), z^n(2), \ldots z^n(N)$, compute $L(z^n(i))$
for $i=1,\ldots, N$, and make a histogram of $L(z^n(i))$. 
We put $(m,n)=(100,200)$ and $N = 10^4$ and 
choose $\epsilon \in\{0.46, 0.5, 0.54, 0.58\}$.
A parity check matrix $A$ is generated once and fixed. 
Fig.~\ref{Distribution_of_Information_Leakage_Given_H} shows 
the histogram of $L(z^n)$. 
We observe that when $\epsilon = 0.54$ which is greater than $100/200=0.5$, 
the average information leakage is small, but 
the conditional information leakage 
takes relatively large value, $6, 7$ and $8$
with probability $0.0081, 0.060$ and $0.044$. 
Thus, from a security perspective, 
the conditional information leakage
is more meaningful than the averaged one.

\subsection{Average distribution of $L(Z^n)$ over randomly generated parity check matrices}
\label{section:average_histogram}

\begin{figure}[t]
\centering 
\includegraphics[height = 50mm]{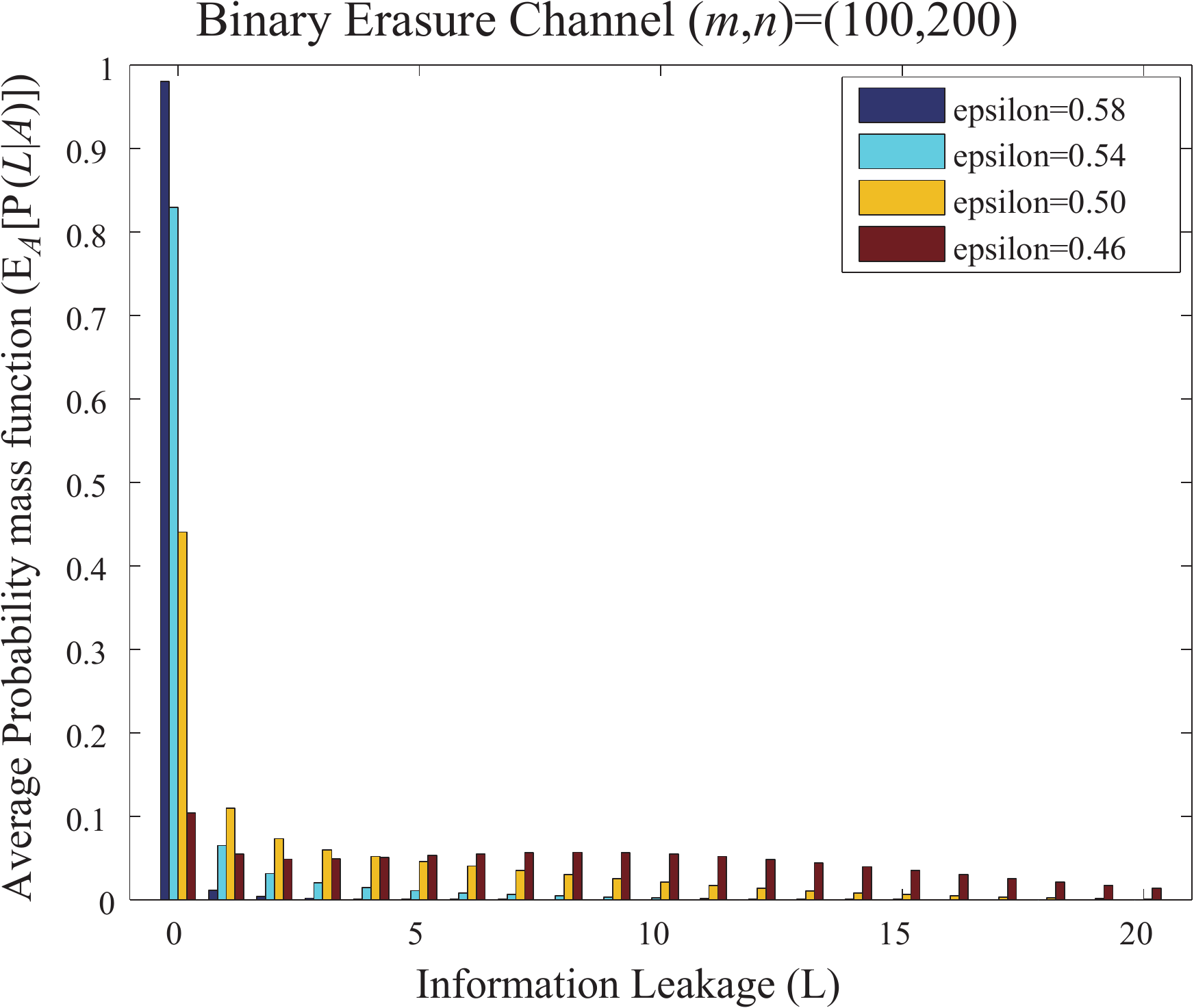}
\caption{The ensemble average of the probability distribution, $\overline{p}_L(\ell)$. }
\label{Average_Distribution_of_Information_Leakage}
\end{figure}

The distribution of $L(Z^n)$, denoted by $p_L(\ell) $, 
depends on the selection of $A$. However,
if $m$ is large enough, $p_L(\ell)$ can be well-approximated by the
ensemble average of $p_L(\ell)$ over all possible $A$, 
as shown in this subsection. 

%
Let us evaluate the ensemble average of
the probability mass function of $L(Z^n)$ over randomly generated parity check matrices.
Suppose Eve's channel is a BEC with erasure probability $\epsilon$. 
To clarify the dependency of the parity check matrix $A$, 
denote the conditional information leakage $L(z^n)$ as $L(z^n|A)$. 
The probability mass function of $L(Z^n| A )$ is defined by
\begin{align}
 p_{L}( \ell ) = \mathrm{Pr} ( L(Z^n | A ) = \ell). 
\label{def:p_L}
\end{align}
Suppose $A$ is generated equally randomly over $\mathbb{F}_2^{m \times n}$.
Let us denote the random variable for $A$ by $\bm{A}$. 
Express the right hand side of (\ref{def:p_L}) by $P_{L|\bm{A}}(\ell|\bm{A})$
We define the average probability distribution of $L(Z^n|A)$ over $\bm{A}$ as 
\begin{align}
 \overline{p}_L(\ell) = {\rm E}_{\bm{A}} [ P_{L|\bm{A}}(\ell|\bm{A})]. 
\label{def:p_L_bar}
\end{align}

We can evaluate $\overline{p}_{L}(\ell)$ by counting the number of matrices in 
$\mathbb{F}_2^{m \times n}$ whose rank is $r$. 
The following function $F(m, n)$ ($m\geq n$)
expresses the number of full-rank matrice in $\mathbb{F}_2^{m\times n}$ : 
$$
  F(m, n)  = \prod_{i=0}^{n-1} (2^m - 2^i).
$$
Then, the probability for $m \times n$ random binary matrix to have rank $r$ is given by 
\begin{align}
  Q( r | m, n) = \frac{1}{2^{n m}} \frac{F(m, r) \cdot F(n, r)}{F(r,r)} 
 \label{Pmnr}
\end{align}
See~\cite{Ferreira2013} for the derivation. 
We have 
\begin{theorem} \label{Theorem.Distribution}
Let $K$ be a random variable following the binomial distribution
with parameter $n$ and $\epsilon$. 
The average probability mass function $\overline{p}_L(\ell)$ of $ L(Z^n| \bm{A})$ over 
the random matrix $\bm{A}$ is given by 
\begin{align}
\overline{p}_{L}( \ell ) 
&= \mathrm{E}_K [ Q(m-\ell | m, K) ] \notag\\
&= \sum_{k = m - \ell }^n \binom{n}{k} (1-\epsilon)^{n-k} \epsilon^k 
Q( m - \ell | m, k ). 
\label{Eq.Theorem.Distribution}
\end{align}
\end{theorem}

\textit{Proof}: From (\ref{def:L}), (\ref{def:p_L}), 
and (\ref{def:p_L_bar}), we have 
\begin{align}
& \overline{p}_L(\ell) \notag \\
&=
\mathrm{E}_{\bm{A}} [ \mathrm{Pr} ( L( Z^n | \bm{A} ) = \ell ) ] \notag \\
&=
\mathrm{E}_{\bm{A}} [ 
\mathrm{Pr} ( \mathrm{rank}[ \{ \bm{a}_j : Z_j = \mathrm{e} \}]  ) = m - \ell ) ] \notag \\
&=
\sum_{A \in \mathbb{F}_2^{m\times n}}
\mathrm{Pr} ( \bm{A} = A) 
\mathrm{Pr} (Z^n = z^n) \notag\\
&\hspace{1cm}\cdot \1 ( \mathrm{rank}[ \{ \bm{a}_j : z_j = \mathrm{e} \}]  ) = m - \ell ) \notag \\
&=
\mathrm{Pr} (Z^n = z^n) 
Q(m-\ell \,|\, m, \# \{ j : z_j = \mathrm{e} \}) \notag \\
&=
\mathrm{E}_{K} [
Q(m-\ell \,|\, m, K)
]
\end{align}
Since $Q(r|m,n)=0$ if $r>\min\{m,n\}$, we have (\ref{Eq.Theorem.Distribution}).
This completes the proof. \hfill $\square$

Fig.~\ref{Average_Distribution_of_Information_Leakage} shows 
the graph of $\overline{p}_{L}( \ell ) $ for $(m,n)=(100, 200)$. 
We observe that the histogram in Fig.~\ref{Distribution_of_Information_Leakage_Given_H}
is well approximated by the graph in Fig.\ref{Average_Distribution_of_Information_Leakage}.

\section{Conclusion} 
We have defined $L(z^n)= I(S^m; Z^n=z^n)$ as 
the conditional information leakage given Eve's received signal
in Wyner's wiretap channel model and proposed to use it 
as a secrecy criterion. 
The standard definition of information leakage 
$I(S^m; Z^n)$ is the expectation of $L(Z^n)$ over $Z^n$. 
We have investigated the probability distribution of $L(Z^n)$. 

We gave a method for computing $L(z^n)$ efficiently, 
which is a modified version of Zhang et al.'s method~\cite{Zhang,Al-Hassan}.
Although Zhang et al.'s method is only applicable to BSCs,
our method canbe applied to any BI-DMCs. 
Because of the space limitation, we only investigated
the probability distribution of $L(Z^n)$ for the case of BSCs and BECs.
Our proposed method will work better in other DMCs than these two examples. 
The case of other DMCs will be investigated in future. 

\section*{Acknowledgment}
The first author would like to thank Professor Ryutaro Matsumoto 
for suggesting him $q$-binomial coefficients, 
also called Gaussian binomial coefficients.
This suggestion lead to the analysis discussed 
in Section~\ref{section:average_histogram}. 

\renewcommand{\appendixname}{Appendix}
\appendix

\section{Proof of Theorem \ref{eqtheo} }
\label{appendix.A}
In this section, we prove Theorem \ref{eqtheo}. 

{\it Proof of Theorem \ref{eqtheo}}:
We first expand $I(S^m; Z^n) = \mathscr{H}(Z^n) - \mathscr{H}(Z^n| S^m) $. 
The codeword $X^n$ is given by (\ref{Coset_Coding}). 
Although the additive noise in the BSC $V^n$ is not uniform,
$Z^n= X^n \oplus V^n$ follows a uniform distribution on $\mathbb{F}_2^{n}$ 
since $X^n$ is a uniform random vector because of its construction. 
Then, $\mathscr{H}(Z^n) = n$. 
Express the noise vector as $V^n = ( \bm{V}_1^T, \bm{V}_2^T)^T$. 
Then the vector of Eve's received symbols is expressed by 
$Z^n = X^n + V^n$, where 
$Z^n = ( \bm{Z}_1^T, \bm{Z}_2^T)^T$ is given by 
\begin{align}
 \bm{Z}_1 &= S^m + A_2 U^{n-m} + \bm{V}_1, \\
 \bm{Z}_2 &= U^{n-m} + \bm{V}_2.
\end{align}

We have the following chain of equalities: 
\begin{align*}
& \mathscr{H}(Z^n | S^m) = \mathscr{H}( \bm{Z}_1, \bm{Z}_2|S^m)\\
& = \mathscr{H}( S^m + A_2 U^{n-m} + \bm{V}_1, U^{n-m} + \bm{V}_2|S^m)\\
& 
\stackrel{\rm (a)}
= \mathscr{H}( A_2 U^{n-m} + \bm{V}_1, U^{n-m} + \bm{V}_2|S^m) \\
& 
\stackrel{\rm (b)}
= \mathscr{H}( A_2 U^{n-m} + \bm{V}_1, U^{n-m} + \bm{V}_2) \\
& 
= \mathscr{H}( A_2 U^{n-m} + \bm{V}_1| U^{n-m} + \bm{V}_2) + \mathscr{H}(U^{n-m} + \bm{V}_2)\\
& 
\stackrel{\rm (c)}
= \mathscr{H}( A_2 U^{n-m} + \bm{V}_1 + A_2(U^{n-m} + \bm{V}_2)| U^{n-m} + \bm{V}_2) \notag\\
& \quad + \mathscr{H}(U^{n-m} + \bm{V}_2)\\
& = \mathscr{H}( \bm{V}_1 + A_2 \bm{V}_2| U^{n-m} + \bm{V}_2) + \mathscr{H}(U^{n-m} + \bm{V}_2)\\
& \stackrel{\rm (d)}
= \mathscr{H}( \bm{V}_1 + A_2 \bm{V}_2) + \mathscr{H}(U^{n-m} + \bm{V}_2)\\
& \stackrel{\rm (e)}
=  \mathscr{H}( \bm{V}_1 + A_2 \bm{V}_2) + (n-m). 
\end{align*}
Step(a) holds because we can remove $S_m$ from
the random variable of the conditional entropy
since $S_m$ is a given random variable. 
Step(b) follows since $U^n$ is generated
independently of $S^m$. 
Step(c) holds since we can add 
a function of $U^{n-m} + \bm{V}_2$ since it is given as the condition. 
Step(d) follows because $U^{n-m}$ is uniform, 
we do not gain any information on $\bm{V}_2$ by knowing 
$U^{n-m} + \bm{V}_2$. This is confirmed more explicitly by 
the following inequality showing that 
$U^{n-m} + \bm{V}_2$ and $\bm{V}_1+A_2 \bm{V}_2$
are independent: 
\begin{align*}
& I(\bm{V}_1 + A_2 \bm{V}_2 ;  U^{n-m} + \bm{V}_2) \\
&= \mathscr{H}(U^{n-m} + \bm{V}_2) - \mathscr{H}(U^{n-m} + \bm{V}_2 | \bm{V}_1 + A_2 \bm{V}_2) \\
& \leq \mathscr{H}(U^{n-m} + \bm{V}_2) - \mathscr{H}(U^{n-m} + \bm{V}_2 | \bm{V}_1, \bm{V}_2) \\
&= \mathscr{H}(U^{n-m} + \bm{V}_2) - \mathscr{H}(U^{n-m}| \bm{V}_1, \bm{V}_2)\\ 
&= \mathscr{H}(U^{n-m} + \bm{V}_2) - \mathscr{H}(U^{n-m})\\ 
&= (n-m) - (n-m) = 0.
\end{align*}
Lastly, Step(e) follows from that $U^{n-m} + \bm{V}_2$ is uniformly 
distributed on $\mathbb{F}_2^{n-m}$. 

Consequently,  we have
\begin{align*}
I(S^m; Z^n) 
& = \mathscr{H}(Z^n) - \mathscr{H}(Z^n| S^m) \\
&= n - \{ \mathscr{H}( \bm{V}_1 + A_2 \bm{V}_2) + (n-m) \}\\
&= m - \mathscr{H}( \bm{V}_1 + A_2 \bm{V}_2) \\
&= m - \mathscr{H}( A V^n).
\end{align*}
This completes the proof. 
\hfill$\square$

\section{ Proof of Theorem~\ref{Theorem_DMC} }
\label{Appendix:Theorem_DMC}
In this section, a proof of Theorem~\ref{Theorem_DMC} is given. 

{\it Proof of Theorem~\ref{Theorem_DMC}}:
We have 
    \begin{align}
      p(s^m|z^n) &= \sum_{x^n} p(s^m,x^n|z^n)\nonumber\\
                     &= \sum_{x^n} p(x^n|z^n) p(s^m|x^n,z^n)\nonumber\\
                     &= \sum_{x^n} \Phi^n(x^n|z^n) \1 (s^m = A x^n), \label{p(s^m|z^n)}
    \end{align}
where $\1$ denotes the indicator function. 
Because $X^n$ follows uniform distribution and the eavesdropper's channel
is memoryless, we have 
    \begin{align}
      \Phi^n(x^n|z^n) = \prod_{i=1}^n \Phi(x_i|z_i). 
    \end{align}
Then, the following chain of equalities holds for the probability generation function of $p(S^m|Z^n=z^n)$:
    \begin{align}
      &G_{S^m|Z^n=z^n}(t) \nonumber\\
      &= \sum_{s^m \in \mathbb F_2^m} p(s^m|z^n)t^{ s^m}\nonumber\\
      &\stackrel{\rm (a)}{=} \sum_{s^m}\sum_{x^n} \Phi(x^n|z^n) \1 (s^m = A x^n)t^{ s^m }\nonumber\\
      \label{temp6}
      &= \sum_{x^n}\Phi^n(x^n|z^n)t^{ A x^n }\\
      \label{temp4}
      &= \prod_{i=1}^n \left( \Phi(0|z_i) + \Phi(1|z_i)t^{ {\bm{a}_i} }\right), 
    \end{align}
where Step (a) follows from (\ref{p(s^m|z^n)}). \hfill$\square$
By expanding (\ref{temp4}) in a way similar to Eqs.(\ref{GAVn-1}) to (\ref{GAVn-2}),
we obtain (\ref{temp22}), which completes the proof.

\section{Proof of Theorem \ref{Theorem_BEC}}\label{proof_theorem_BEC}
  In this section, we prove Theorem~\ref{Theorem_BEC}.
  To this aim, we give two lemmas.

  \begin{lemma} 
  \label{lemma.1}
  Suppose that the encoder uses a coset code and 
  Eve's channel is a BEC. Then, the information leakage to Eve 
  only depends on the position of the erasure occurred and 
  is independent of whether $z_i = 0 $ or $z_i = 1$ is received. 
  \end{lemma}
  \begin{lemma} 
    \label{lemma.2}
    Suppose the encoder uses a coset code with parity check matrix 
    $A = [\bm{a}_1, \ldots, \bm{a}_n]$ and Eve's channel is a BEC. 
    Assume $x^n = 0^n$ is transmitted. 
    Let $z^n$ be Eve's received signal and let 
    $\mathcal{J}_{ \mathrm{e} }(z^n) = \{ j : z_j = e \}$. 
    Let $w$ be the rank of $ [ \bm{a}_{j} | j \in 
    \mathcal{J}_{ \mathrm{e} }(z^n)]$. 
    Assume $\bm{a}_{j_1}$,  $\bm{a}_{j_2}$\ldots $\bm{a}_{j_w}$ are independent. 
    Then, we have 
    \begin{align} 
       G_{S^m|Z^n=z^n} (t) = \frac{1}{2^w} \prod_{ i = 1}^{w} 
       \left(1 + t^{ \bm{a}_{j_i} }  \right)
       \label{Eq.lemma.2}
    \end{align}
  \end{lemma}

\textit{Proof of Lemma~\ref{lemma.1}}:
Define the index set 
$\mathcal{J}_{\rm e}(z^n) = \{ j | z_j = \mathrm{e} \}$ 
and $\mathcal{J}_{1}(z^n) = \{ j | z_j = {1} \}$.
Then, we have
\begin{align}
G_{S^m|Z^n=z^n} (t)
&= 
\prod_{i \in \mathcal{J}_1(z^n) } t^{ \bm{a}_i } 
\prod_{j \in \mathcal{J}_{\rm e}(z^n) } 
\left(
\frac12 +
\frac12 
t^{ \bm{a}_j } 
\right) \notag \\
&= 
t^{ \sum_{i \in \mathcal{J}_1 (z^n) } \bm{a}_i } 
\prod_{j \in \mathcal{J}_{\rm e} (z^n) } 
\left(
\frac12 +
\frac12 
t^{ \bm{a}_j } 
\right)
\end{align}
Define
\begin{align}
\hat G_{S^m|Z^n=z^n} (t)
= 
\prod_{j \in \mathcal{J}_{\rm e} (z^n)} 
\left(
\frac12 +
\frac12 
t^{ \bm{a}_j } 
\right).
\end{align}
Then $\hat G_{S^m|Z^n=z^n}(t)$ is the probability generating function 
if all $z_i = 1$ in $z^n$ is replaced by $0$.
Put its expansion as
$ \hat G_{S^m|Z^n=z^n} (t)
= 
\sum_{ \bm{s} \in \mathbb{F}_2^m } \hat{\beta}_{\bm{s}} t^{\bm{s}} 
$. 
Then we have
\begin{align}
 G_{S^m|Z^n=z^n} (t)
& =
\sum_{ \bm{s} \in \mathbb{F}_2^m } \hat \beta_{\bm{s}}  
t^{ \bm{s} \oplus \sum_{ i \in \mathcal{J}_1 } \bm{a}_i } \notag \\
& =
\sum_{ \bm{s} \in \mathbb{F}_2^m } 
     \hat \beta_{ \bm{s} \oplus \sum_{i \in \mathcal{J}_1 (z^n) } \bm{a}_i } 
t^{ \bm{s}  },
\end{align}
which implies $\beta_{ \bm{s} } = \hat \beta_{ \bm{s} \oplus 
\sum_{i \in \mathcal{J}_1 (z^n)} \bm{a}_i } $
and thus the distribution of $S^m$ given $z^n$ is 
a permutation of the distribution computed from $\hat G_{S^m|Z^n=z^n}(t)$. 
Since entropy does not change by a 
permutation of the probability distribution, 
we have Lemma~\ref{lemma.1}. \hfill $\square$

  \textit{ Proof of Lemma~\ref{lemma.2} }
  Since $\bm{a}_{j_1},\bm{a}_{j_2},...\bm{a}_{j_w}$ are linearly independent, 
  we can express ${\bm{a}_{j_k}}$
  for $k \in \{ w+1,w+2, \ldots, | \mathcal{J}_{\rm e} (z^n)| \}$ as 
  \begin{align*}
    {\bm{a}_{j_k}} = \sum_{i=1}^w d_{k,i}{\bm{a}_{j_i}} \ \ \ (d_{k,i}\in \mathbb F_2\ {\rm for\ }i=1, 2, \ldots, w)
  \end{align*}
  for some coefficient $d_k,i$. By Lemma \ref{lemma.1}, we can assume $x^n=0^n$. 
  Substituting the conditional probability of the backward channel $\Phi(0|z_i)$ into (\ref{temp4}) gives 
  \begin{align*}
    &G_{S^m|Z^n = z^n}(t) \nonumber\\
    &= \prod_{l=1}^w \left(\frac12 + \frac12 t^{ {\bm{a}_{j_l}}}\right) \prod_{k=w+1}^{| \mathcal{J}_{\rm e}(z^n)|}\left(\frac12 + \frac12 t^{ {\bm{a}_{j_k}} }\right)\nonumber\\
    &= \frac1{2^{| \mathcal{J}_{\rm e}(z^n)|}}\prod_{l=1}^w \left(1 +  t^{ {\bm{a}_{j_l}} }\right)\prod_{k=w+1}^{| \mathcal{J}_{\rm e}(z^n)|}\left(1 +  t^{ \sum_{i=1}^wd_{k,i}{\bm{a}_{j_i}}}\right)\nonumber\\
    &= \frac1{2^{| \mathcal{J}_{\rm e}(z^n)|}}\prod_{l=1}^w \left(1 +  t^{ {\bm{a}_{j_l}} }\right)
                             \left(1 +  \prod_{i=1}^w t^{ d_{w+1,i}{\bm{a}_{j_i}} }\right)\nonumber\\
    &\hspace{3cm}\prod_{k=w+2}^{| \mathcal{J}_{\rm e}(z^n)|}\left(1 +  t^{ \sum_{i=1}^wd_{k,i}{\bm{a}_{j_i} } }\right)\nonumber\\
    &= \frac1{2^{| \mathcal{J}_{\rm e}(z^n)|}}\Bigg\{ \prod_{i=1}^w\left(t^{  d_{w+1,i}{\bm{a}_{j_i}} } +  t^{ (1\oplus d_{w+1,i}) {\bm{a}_{j_i}}  }\right)\nonumber\\
    &+ \prod_{i=1}^w\left(1 +  t^{ {\bm{a}_{j_i}}  }\right)\Bigg\}\prod_{k=w+2}^{|\mathcal{J}_{\rm e}(z^n)|}\left(1 +  t^{ \sum_{i=1}^wd_{k,i}{\bm{a}_{j_i}}  }\right). 
  \end{align*}
  Because 
  $t^{ d_{w+1,i}{\bm{a}_{j_i}}  } +  t^{ (1\oplus d_{w+1,i}) {\bm{a}_{j_i}}  } 
   = 1 +  t^{ \bm{a}_{j_i}  }$
  holds for $i\in \{1,2,...,w\}$, we have 
  \begin{align*}
    G_{S^m|Z^n = z^n}(t) &= \frac1{2^{| \mathcal{J}_{\rm e}(z^n)|-1}}\prod_{i=1}^w\left(1 +  t^{ {\bm{a}_{j_i}} }\right)\nonumber\\
    &\quad \times \prod_{k=w+2}^{|\mathcal{J}_{\rm e}(z^n)|}\left(1 +  t^{ \sum_{i=1}^wd_{k,i}{\bm{a}_{j_i}}}\right). 
  \end{align*}
  Continuing the same procedure for $k= w+2,...,| \mathcal{J}_{\rm e}(z^n)|$, we obtain Lemma \ref{lemma.2} \hfill $\square$

  \textit{ Proof of Theorem~\ref{Theorem_BEC} }
  As Lemma~\ref{lemma.2}, let $w$ be the rank of $[ \bm{a}_{j} | j \in 
  \mathcal{J}_{\mathrm{e}}(z^n)]$
  and assume $\bm{a}_{j_1},\bm{a}_{j_2},...\bm{a}_{j_w}$ are linearly independent. 
  Then, for all $\sum_{i=1}^w b_i \bm{a}_{j_i}$, $b_i \in \mathbb{F}_2$ are different.
  By expanding (\ref{Eq.lemma.2}), we have
  $ \beta_{ \bm{s} } = \mathrm{Pr}( S^m = \bm{s} ) = \frac{1}{2^w}$ if 
  $ \bm{s} = 
  \sum_{i=1}^w b_i \bm{a}_{j_i}$ for some $b_i$s and $\beta_{ \bm{s} } =0$ otherwise. 
  This completes the proof. \hfill$\square$

\section{Proof of Theorem~\ref{Theorem_conditional} }
\label{Appendix:Theorem_conditional}
This section gives a proof of Theorem~\ref{Theorem_conditional}. 

\textit{Proof of Theorem~\ref{Theorem_conditional}:} 
We have the following equality. 
  \begin{align*}
     \mathscr{H}(S^m|Z^n = z^n) 
  &= \mathscr{H}(S^m \oplus H Z^n |Z^n = z^n) \\
  &= \mathscr{H}(H V^n |Z^n = z^n) \\
  \end{align*} 
However, we have
\begin{align}
& p_{HV^n|Z^n}(y^m|z^n) \notag \\
& = \frac{1}{p_{Z^n}(z^n)} \sum_{x^n} \sum_{v^n}
p_{X^n V^n HV^n Z^n} (x^n, v^n, y^m, z^n)\notag \\
& = \frac{1}{p_{Z^n}(z^n)} \sum_{x^n} \sum_{v^n}
p_{X^n}(x^n) 
p_{V^n}(v^n) 
\1(y^m = H v^n) \notag\\ 
& \hspace{2cm} \cdot \1 (z^n = x^n \oplus v^n)\notag \\
& = \frac{1}{p_{Z^n}(z^n)} \sum_{x^n} 
p_{X^n}(x^n) 
p_{V^n}(x^n \oplus z^n) \notag\\
&\hspace{2cm} \cdot \1(y^m = H (x^n \oplus z^n) ) \notag \\
& = \frac{1}{p_{Z^n}(z^n)} \sum_{x^n} 
p_{X^n}(x^n \oplus z^n) 
p_{V^n}(x^n) 
\1(y^m = H (x^n) )
\label{pHVngivenZn}
\end{align}
Since both $X^n$ and $Z^n$ follow the uniform distribution,
Eq.(\ref{pHVngivenZn}) shows  that $p_{HV^n|Z^n}(y^m|z^n)$ is independent of 
$z^n$. This completes the proof. \hfill $\square$



\end{document}